\documentclass[prl,showpacs,preprintnumbers,amsmath,amssymb]{revtex4}

\usepackage[dvips]{graphics}
\usepackage{tikz}

\begin{document}

\title{A Microscopic Quantum Model For the Experiment Coupling Qubits to a Tardigrade}

\author{Vlatko Vedral}
\affiliation{Clarendon Laboratory, University of Oxford, Parks Road, Oxford OX1 3PU, United Kingdom and\\Centre for Quantum Technologies, National University of Singapore, 3 Science Drive 2, Singapore 117543 and\\
Department of Physics, National University of Singapore, 2 Science Drive 3, Singapore 117542}

\date{\today}

\begin{abstract}
\noindent We provide a quantum model for the recent experiment coupling a tardigrade to superconducting qubits. A number of different perspectives are discussed with the emphasis placed on quantum entanglement between different subsystems involved in the description.  
\end{abstract}


\maketitle                           

A recent experiment with which I was involved \cite{Rainer} has generated some controversy regarding its interpretation and this is what I would like to address in this brief note. The question, as frequently is the case, ultimetely boils down to understanding the notion of classicality within an otherwise fully quantum mechanical universe. The account will be completely non-relativistic since the effects of retardation (no pun intended) are completely irrelevant. 

The first part of the experiement was to couple a tardigrade to a single superconducting qubit. The result of this coupling (the tardigrade was inserted between the capacitor plates of the qubit) was the frequency shift between the logical states of the qubit, which was experimentally observed (through microwave spectroscopy whose details are irrelevant for our discussion). The presence of the tardigrade therefore alters the resonant frequency of the qubit (with respect to what it was measured to be without the tardigrade). 

The key issue to explain is the physical origin of this shift. The specific question is: was the tardigrade actually entangled to the qubit? In order to address this question, we would like to claim the following: the shift can be viewed as either an instance of the Lamb shift, or the Stark shift, or the Casimir effect, or, indeed, as a manifestation of the van der Waals force between the qubit and the tardigrade. All four are genuine quantum effects of the zero-point energy (involving entanglements of different kinds, as will be explained below) and are really just different ways of partitioning one and the same Hamiltonian that ultimately fully accounts for it. All these perspectives, accounting for the resulting energy shift of a dipole immersed in a medium of different dipoles, are well-known. A thorough and insightful discussion can, for instance, be found in the beautiful book by Milonni \cite{Milonni} (which is sadly too long to be put on Twitter, the medium where all serious debates, scientific or otherwise, are currently being held).

We will now make the following assumptions. The tardigrade itself is modelled as a dielectric medium consisting of many microscopic quantum dipoles (needles to say, the experiment was not able to tell exactly which molecules in the tardigrade responded). All of these dipoles couple to the underlying (quantum) electromagnetic field. The qubit itself also couples to the underlying electromagnetic field. The total Hamiltonian (of the qubit plus the tardigrade plus the electromagnetic field) is given by:
\begin{equation}
H = \frac{\hbar \omega}{2}\sigma_z + \sum_n \hbar \omega_n a^{\dagger}_n a_n + \sum_m \hbar \Omega_m b^{\dagger}_m a_m + \sigma_x \sum_k \hbar g_k (a^{\dagger}_n + a_n) + \sum_l \hbar f_l (b^{\dagger}_l + b_l)(a^{\dagger}_l + a_l)\; ,
\end{equation}
where the first term is the energy of the qubit, the second is the energy of the surrounding electromagnetic field, the third is the energy of the dipoles inside the tardigrade. The fourth and fifth terms are the intereaction terms, which we will call $H_I$, and which are key to our discussion, the former being the coupling between the qubit and the electromagnetic field and the latter representing the coupling between the tardigrade dipoles and the electromagnetic field. Note that we have not taken into account the fact that the experiment involves an electromagnetic driving field which excites the qubit in order to register its emission. We have also assumed that the dipoles are undamped. This can be modified by adding the phononic degrees of freedom which couple to the dipoles in order to generate damping (ultimately resulting in the line broadening of the dipole transitions). Neither of the two effects, the driving and the damping, are directly relevant for the points that follow. 

As far as the calculation of the shift of energies of the qubit, the easiest way to proceed is to use the second order perturbation theory. We assume that the tardigrade dipoles are initially in their respective ground state (as we are indeed at miliKelvin temperatures) and the same is true for the state of the electromagnetic field (it is in the quantum vacuum). We are interested in the energy shifts of the qubit due to the coupling to the electromagnetic vacuum which in turn is coupled to the tardigrade dipoles. We will use the following notation: $|\bar i\rangle = |i\rangle|0\rangle|0\rangle$, where the first ket belongs to the qubit ($i=0,1$), the second ket represents the vacuum state of the field (i.e. each mode being in the ground state) and the third ket is the ground state of all of the tardigrade dipoles. Then the well-known perturbation formula implies:
\begin{equation}
\Delta E_i = \sum_{n \neq i} \frac{\langle \bar i|H_I|n\rangle\langle n H_I|\bar i\rangle}{\hbar \omega_i - \hbar \omega_n}\; ,
\end{equation}
which consists of the contributions of both of the interactions terms. We will evaluate this expression in a somewhat simplified scenario below just for the sake of giving a concrete illustration. What is ultimately accessible to the experiment (among other things that are not directly relevant for the discussion) is the difference $E_1+\Delta E_1 - (E_0+\Delta E_0)$ which can be contrasted with the qubit energy levels without the tardigrade (i.e. $E_1- E_0$). 

Now, focussing on the physical interpretations of this formula, the story can be told in a number of different ways. First, we can think of the dipoles of the tardigrade modifying the electromagnetic vacuum through their interaction. This is the same as a dielectric's modification of the underlying refractive index which increases as result. One can then work directly with the quantum electromagnetic potential operators in the dielectric \cite{Milonni2} and couple these directly to the qubit. The qubit levels will thus be shifted. This could be thought of as the Lamb shift in the presence of the dielectric.  In this picture, the entanglement between the tardigrade dipoles and the electromagnetic field is the reason for the increased refractive index. On the other hand, the entanglement between this now dressed field and the qubit is what gives us the observed energy shifts. Clearly, entanglements matter. 

The interpretation number two is to say that the tardigrade-dressed electromagnetic field fluctuates. The energy fluctuations of this dressed field then disturb the qubit which couples to it through $\alpha \langle \bar E\rangle$, where $\alpha$ is the resulting dielectric permeability (which itself is a function of the refractive index). The energy shift observed in the qubit is then seen as the Stark shift due to the dressed vacuum fluctuations. At this level of ``coarse graining" one does not need further entanglement to explain anything. Namely, if the dressed fluctuations are taken as the fundamental quantity then they could be just thought of as providing the ``classical" force that couples to the qubit to shift its levels. However, as is well know \cite{Milonni}, the vacuum fluctuations are fundamentally quantum in nature and cannot ever be accounted for classically (despite many attempts by many people).  

The third interpretation simply rests on the fact that it is the zero point energy of the electromagnetic field that becomes altered by the presence of the qubit and the tardigrade dipoles. The resulting change in the energy of the vacuum is, of course, the Casimir effect, which is what is observed in the qubit energy shifts. This is still described by the same Hamiltonian above, it is just that the order in which we take different terms into account changes when we change our interpretation. This interpretation resonates with the quantum-field-theoretic view in which all the relevant quantities are contained in the vacuum persistence amplitude, namely the amplitude that the old vacuum will remain the same after it has been perturbed by something (like, for instance, a qubit or a tardigrade). 

Needless to say, the tardigrade experiment did not require the full mathematical treatment of the above kind. That would have been an overkill. The simple reason is that, in the first part of the experiment, the only accessible quantity was the frequency shift (and none of the coupling constants in the Hamiltonian or the tardigrade dipole frequencies were known or even accessible in this experiment) and this could easily be accouted for by a far simpler model of a qubit coupled to a quantized simple harmonic oscillator (approximately representing the collective state of all the dipoles pertaining to the tardigrade - see \cite{Vedral} for a similar treatment of another experiement involving a living system strongly coupled to light). In fact, a direct modulation of the qubit frequency ($\omega \rightarrow \omega/n$, where $n$ is the refractive index due to the presence of the tardigrade) clearly suffices to account just for the frequency shift (but it fails to explain the relevant ``physics"). 

Here we reach the fourth interpretation of the qubit energy shifts and the treatment here will very much resembles the London account of the van der Waals forces \cite{London}. The logic is as follows. We take both the qubit and the tardigrade dipoles as already dressed by the electromagnetic field. In this way, the field is eliminated from the Hamiltonian. This is probably the most sensible account since the frequencies of the qubit and the dipoles are never really ``naked" but are always measured when bathed in the surrounding electromagnetic vaccum. We then assume that the dipoles in the tardigrade couple directly (through the induced electrostatic forces) to the qubit. As we said it is perfectly fine to view this interaction as instantaneous (even though it is not in fact) since the relativisitic corrections are indeed negligible (the so-called Casimir-Polder forces \cite{Landau}). 

As far as entanglement is concerned, the entanglement between the field and the tardigrade as well as between the qubit and the field are automatically included so they are not explicit in the remaining model. The only relevant entanglement is between the qubit and the tardigrade. We will assume the effective Hamiltonian to be (see e.g. \cite{Holstein}):
\begin{equation}
H_{eff} = \frac{p^2_1}{2m} + \frac{m\omega_0^2x_1^2}{2}+\frac{p^2_2}{2m} + \frac{m\omega_0^2x_2^2}{2} + \frac{e^2}{4\pi\epsilon_0}\left(\frac{1}{R}+\frac{1}{R+x_1-x_2}+\frac{1}{R+x_1}+\frac{1}{R-x_2}\right)
\end{equation}
where we have two free harmonic oscillator hamiltonian's (for simplicity assumed to be the same, but this does not alter the relevant conclusions) as well as the last term representing their interaction (again I have overslimplified by assuming that each has just one electron worth of charge). I have deliberately tried to make this look as ``classical" as possible by stripping away all other ineractions. Even though the qubit is represented as a harmonic oscillator (which it is as it resembles an LC circuit), only the lowest two energy levels will be experimentally relevant (from that perspective it is also immaterial that the superconducting qubit is actually anharmonic). The interaction term is simply just the dipole-dipole coupling ($\approx e^2 x_1x_2/R^3$). 

This system can be diagonalised exactly, but we do not lose much by using 
the second order perturbation which gives us the energy shift in the qubit states. For instance, the shift in the ground state energy is 
\begin{equation}
\Delta E_0 =  \left(\frac{e^2}{2\pi\epsilon_0 R^3}\right)^2 \frac{\langle 1,1|x_1x_2|0,0\rangle}{-2\omega_0} = -\frac{e^4}{32\pi^2\epsilon_0^2 m^2 \omega_0^3 R^6} \; ,
\end{equation}
and we obtain the usual van der Waals $R^{-6}$ dependence. The minus sign indicates that the force is attractive and this is the quantum explanation for how neutral molecules can still attract one another. Quantum physics allowes the energy to be lower than that of the individual harmonic oscillators and this is because, even at zero termperatures, the positions and momenta of the oscillator still have an uncertainty associate with them (again, these uncertainties are just the zero-point fluctuations arising from the non-commutativity of the respective positions and momenta). 

We can equally well compute the shift in the excited state. Thinking of the experiment this way, one could say that the qubit and the tardigrade actually form a complex hybrid molecule through the van der Waals attraction (it is, to the best of my knowledge, the first molecule composed on a living and an inanimate object). The qubit and the tardigrade, when viewed this way, are certainly quantum entangled, albeit not necessarily maximally. The degree of entanglement depends on the strength of the coupling between them \cite{Vedral}.  It is relatively easy to compute the ground state of the system exactly and therefore calculate the amount of entanglement, but I refrain from doing so since there is a vast amount of literature on this topic. 

Instead, it is worth mentioning Schwinger's account of the Casimir effect \cite{Schwinger} because it also ``factors out" the electromagnetic field and bases the whole calculation on just the relevant sources (that is the whole point of Schwinger's approach to quantum electrodynamics). As such it is similar to the van der Waals picture where the sources couple directly to one another while the field it taken into account ``clasically" (in the form of the relevant propagators in Schwinger's case).  Note also that for the simple harmonic oscillator the polarizability is given by $\alpha = 2e^2/m\omega_0^2$, which shows us why the van der Waals effect can also be interpreted as the Stark shift due to zero point field (as elaborated on in \cite{Holstein}). 

Finally, it might be important to stress one more aspect of the experiement. The tardigrade experiment actually went further than just the frequency shift measurements. Firstly, different quantum states of the qubit-tardigrade hybrid system were prepared and tomographically confirmed \cite{Rainer}. Secondly, and more importantly, the experiement coupled another superconducting qubit to the first qubit with the tardigrade. The full quantum state tomography was then performed to show that this engineered state between the two qubits (one alone and the other ``dressed" with a tardigrade) was maximally entangled. 

But what does this prove? Other than an exceptionally high level of experimental sophistication, it proves that each of the qubits is actually a bone fide quantum system. Not only can each be prepared in a general superposition, but they could also be fully entangled to one another. This means that whatever the tardigrade has done to the qubit it sits on (be it the Stark shift, the Lamb shift, the Casimir effect or the van der Waals attraction) is not incoherent and detrimental to its quantum behaviour. 

Returning to the initial question, does this mean that the qubit and the tardigrade have been entangled? Both models above (the full and the effective) would suggest so. But (and there is always the same kind of a ``but" in these discussions) to conclusively verify this, one would have to be able to measure the tardigrade independently from the qubit it is modulating. It would suffice to violate Bell's inequalities by measuring two complementary observables on the tardigrade alone and two other complementary observables on the qubit alone (as already indicated in \cite{Vedral}). This, unfortunately, is still beyond the current experimental capabilities. However, it is certainly something to aim for and, who's to say, it might even be achieved sooner rather than later given the tremendous developments in quantum technologies around the world.

\textit{Acknowledgments}: The author thanks Rainer Dumke, Chiara Marletto and Tomek Paterek and for many discussions related to this topic. His research is supported by the National Research Foundation and the Ministry of Education in Singapore and administered by Centre for Quantum Technologies, National University of Singapore.

\end{document}